\documentclass[preprint,amsmath,amssymb,aps,prl,superscriptaddress,showpacs]{revtex4-1}
\usepackage[latin9]{inputenc}
\usepackage{amsmath}
\usepackage{esint}

\makeatletter
\usepackage{graphicx}% Include figure files
\usepackage{dcolumn}% Align table columns on decimal point
\usepackage{bm}% bold math
\makeatother

\begin{document}

\title{Photons, phonons, and plasmons with orbital angular momentum in plasmas}

\author{Qiang Chen}

\thanks{cq0405@ustc.edu.cn}

\affiliation{Luoyang Electronic Equipment Testing Center, Luoyang 471000, China }

\affiliation{School of Nuclear Science and Technology and Department of Modern
Physics, University of Science and Technology of China, Hefei, Anhui
230026, China}

\author{Hong Qin }

\thanks{hongqin@ustc.edu.cn}

\affiliation{School of Nuclear Science and Technology and Department of Modern
Physics, University of Science and Technology of China, Hefei, Anhui
230026, China}

\affiliation{Plasma Physics Laboratory, Princeton University, Princeton, NJ 08543,
USA}

\author{Jian Liu}

\affiliation{School of Nuclear Science and Technology and Department of Modern
Physics, University of Science and Technology of China, Hefei, Anhui
230026, China}

\affiliation{Key Laboratory of Geospace Environment, CAS, Hefei, Anhui 230026,
China}
\begin{abstract}
Exact eigen modes with orbital angular momentum (OAM) in the complex
media of unmagnetized homogeneous plasma are studied. Three exact
eigen modes with OAM are discovered, i.e., photons, phonons, and plasmons.
It is found that an OAM photon can be excited by two familiar Bessel
modes without OAM. For the phonons and plasmons, their OAM are carried
by the electrons and ions. The OAM modes in plasmas and their characteristics
can be explored for various potential applications in plasma physics
and accelerator physics.
\end{abstract}

\pacs{52.35.Hr, 52.35.Fp, 71.45.Gm, 52.35.We, 42.50.Tx, 52.50.Sw}

\maketitle
During the last quarter century, the generation, transmission, conversion
and detection techniques of photon orbital angular momentum (OAM)
experienced significant advances, due to its wide applications in
quantum information, particle manipulation, non-classical imaging,
nanotechnology and even astronomy \cite{r1,r2,r3,r4,r5,r6,r7,r8,r9,r10,r11,r12,r13,r14}.
In 1990, Tamm \emph{et al.} generated Laguerre-Gaussian (LG) mode
laser beams which have helical wave fronts and can drive neutral atoms
and molecules \cite{r1}. Allen \emph{et al.} first demonstrated that
light beams with an azimuthal phase distribution carries an angular
momentum independent of the polarization photon state \cite{r2}.
The lost part of photon angular momentum embedded in twisted electromagnetic
beam or optical vortex was found. Recently, the extension of photon
OAM technology from visible to radio frequency (RF) leads to more
potential scientific and engineering applications \cite{r7}. All
previous studies are based on paraxial optics with slow varying envelope
approximation in vacuum or crystal, and exact solutions of photons
with OAM in complex media are seldom mentioned.

Electromagnetic waves in plasmas and their interaction with charged
particles play a crucial role in plasma physics and accelerator physics.
RF waves are employed to accelerate particles in modern accelerators
\cite{Davidson01-all}, and to heat plasmas and drive current in magnetic
fusion devices \cite{Fisch87}. They are also the most effective plasma
diagnostic tools. The coupling from injection waves to fusion plasmas
can excite abundant eigen modes, such as the electron cyclotron wave,
ion cyclotron wave and Bernstein wave. Different modes have different
propagation properties, such as the accessibility and absorption characteristics,
which determine their applicability. Although this classical problem
has been intensively studied with wide applications, no attention
has ever been paid to the OAM carried by these waves in plasmas.

Does a plasma support electromagnetic or other type of eigen modes
with OAM? If so, can they be utilized to achieve better diagnostics,
heating, current drive, and particle acceleration? In the present
study, we address these two questions. In the past, measurements of
the interaction between a RF wave with OAM and plasma vortex was made,
and some theoretical descriptions based on the LG mode were given
by Mendonca, Thide \emph{et al.} \cite{r20,r21,r22}. However, LG
modes are solutions under the scalar paraxial approximation assuming
a slowly varying envelope. They are not rigorous solutions of the
vector Maxwell equations. Especially, in the complex media of plasmas,
detailed and careful analysis should be performed on the vector Maxwell
equation with a proper self-consistent model for plasmas. In this
paper, we adopt a two-fluid system which self-consistently couples
the dynamics of electrons and ions with the vector Maxwell equations.
We describe, for the first time, three classes of rigorous solutions
of the system that can be identified as photons, phonons, and plasmons
with OAM. They correspond to the electromagnetic, ion acoustic, and
Langmuir waves in plasmas. The OAM eigen modes in plasmas have azimuthal
phase distribution and Bessel-type radial dependency, but are different
from normal Bessel modes without OAM. Very interestingly, one electromagnetic
OAM eigen mode can be excited by two normal Bessel modes without OAM.
The OAM of different plasma components are closely related to the
charge polarities. For phonons and plasmons, all OAM are carried by
the electrons and ions. Based on their OAM spectrum modulation, power
concentration structure and rotation properties, the OAM eigen modes
have important potential applications in plasma diagnostics, heating,
current drive in magnetic fusion devices and driving rotating charged
particle beams with enhanced stability in high-intensity accelerators.

To study the small amplitude electromagnetic waves in a plasma, we
start from a linearized two-fluid system with self-consistent electromagnetic
field determined by the Maxwell equations, 
\begin{eqnarray}
\frac{\partial}{\partial t}n_{1\alpha}+\bigtriangledown\cdot n_{0\alpha}\mathbf{V_{\alpha}} & = & 0,\label{eq:1}\\
\frac{\partial}{\partial t}\mathbf{V_{\alpha}} & = & \frac{q_{\alpha}}{m_{\alpha}}\mathbf{E},\label{eq:2}\\
\bigtriangledown\times\mathbf{E} & = & -\frac{\partial}{\partial t}\mathbf{B},\label{eq:3}\\
\bigtriangledown\times\mathbf{B}-\frac{1}{c^{2}}\frac{\partial}{\partial t}\mathbf{E} & = & \mu_{0}\mathbf{J}=\mu_{0}\sum_{\alpha=e,i}q_{\alpha}n_{0\alpha}\mathbf{V}_{\alpha},\label{eq:4}\\
\bigtriangledown\cdot\mathbf{E} & = & \frac{q}{\varepsilon_{0}}=\frac{1}{\varepsilon_{0}}\sum_{\alpha=e,i}q_{\alpha}n_{1\alpha},\label{eq:5}\\
\bigtriangledown\cdot\mathbf{B} & = & 0,\label{eq:6}
\end{eqnarray}
where the subscript $\alpha$ denotes electron (e) or ion (i) component,
$\left(\mathbf{E},\,\mathbf{B},\,\mathbf{V}_{\alpha},\,n_{1\alpha}\right)$
represent the first order perturbed fields, $\,n_{0\alpha}$ is the
equilibrium density, and other variables have their usual meanings.
The equilibrium is assumed to be cold, homogeneous, and unmagnetized
and satisfies the neutrality condition $\sum_{\alpha=e,i}q_{\alpha}n_{0\alpha}=0$.
The thermal effects will be considered in the second half of the paper.

The linear system \eqref{eq:1}-\eqref{eq:6} admits two approximations.
The first is the electromagnetic approximation where the quasi-neutrality
condition $\sum_{\alpha=e,i}q_{\alpha}n_{1\alpha}=0$ is assumed and
the system is reduced to 
\begin{eqnarray}
c^{2}\bigtriangledown^{2}\mathbf{E} & = & \frac{\partial^{2}\mathbf{E}}{\partial t^{2}}+\omega_{p}^{2}\mathbf{E},\label{eq:7}\\
\bigtriangledown\cdot\mathbf{E} & = & 0,\label{eq:8}
\end{eqnarray}
where $\omega_{p}=\sqrt{\omega_{pe}^{2}+\omega_{pi}^{2}}$ and $\omega_{p\alpha}=\sqrt{q_{\alpha}^{2}n_{0\alpha}/\varepsilon_{0}m_{\alpha}}$.
The second is the electrostatic approximation where the perturbed
magnetic field is negligible and the system is reduced to 
\begin{eqnarray}
\frac{\partial^{2}\mathbf{E}}{\partial t^{2}}+\omega_{p}^{2}\mathbf{E} & = & 0,\label{eq:9}\\
\bigtriangledown\times\mathbf{E} & = & 0.\label{eq:10}
\end{eqnarray}
It turns out that a monochromatic mode with OAM can be constructed
in the cylindrical coordinates with azimuthal phase distribution $\mathrm{e}^{il\phi}$,
where the integer $l$ is the azimuthal mode number. In quantum optics,
it is known as topological charge describing the degree of phase helicity
\cite{r10}. The mode assumes the form of 
\begin{eqnarray}
\mathbf{E}=\sum_{\beta}E_{\beta}\left(r,\phi,z\right)\mathbf{e}_{\beta}=\sum_{\beta}E_{\beta}\left(r\right)\mathrm{e}^{i\left(l\phi+kz-\omega t\right)}\mathbf{e}_{\beta},\label{eq:13}
\end{eqnarray}
where the subscript $\beta=r,\,\phi$, or $z$ denotes cylindrical
coordinates, and $E_{\beta}(r)$ is a function of the radial coordinate.
The $z-$direction is the space-time averaged propagation axis.

We first look at the electromagnetic modes, i.e., photons, with OAM.
Substituting Eq.~(\ref{eq:13}) into Eq.~(\ref{eq:7}), we obtain
the eigen equation of the electromagnetic modes, 
\begin{eqnarray}
\left(\begin{array}{ccc}
S & -iD & 0\\
iD & S & 0\\
0 & 0 & P
\end{array}\right)\left(\begin{array}{c}
E_{r}(r)\\
E_{\phi}(r)\\
E_{z}(r)
\end{array}\right)=0,\label{eq:14}
\end{eqnarray}
where $D=2l/r^{2}$, and other matrix elements are defined as, 
\begin{eqnarray}
S & = & \frac{1}{r}\frac{\partial}{\partial r}\left(r\frac{\partial}{\partial r}\right)-\left(\frac{l^{2}+1}{r^{2}}+k^{2}\right)+\frac{\omega^{2}-\omega_{p}^{2}}{c^{2}},\label{eq:15}\\
P & = & \frac{1}{r}\frac{\partial}{\partial r}\left(r\frac{\partial}{\partial r}\right)-\left(\frac{l^{2}}{r^{2}}+k^{2}\right)+\frac{\omega^{2}-\omega_{p}^{2}}{c^{2}}.\label{eq:16}
\end{eqnarray}
In terms of $E^{\pm}(r)\equiv E_{r}(r)\pm iE_{\phi}(r)$, Eq.~(\ref{eq:14})
can be rewritten as 
\begin{eqnarray}
\left(\begin{array}{ccc}
S-D & 0 & 0\\
0 & S+D & 0\\
0 & 0 & P
\end{array}\right)\left(\begin{array}{c}
E^{+}(r)\\
E^{-}(r)\\
E_{z}(r)
\end{array}\right)=0,\label{eq:17}
\end{eqnarray}
which shows that the $E^{+}(r)$, $E^{-}(r)$, and $E_{z}(r)$ components
are decoupled. A class of special solutions of eigen equation which
satisfy the finite boundary condition at axis are $E^{+}(r)=\widetilde{E^{+}}J_{l+1}\left(\mu r\right)$,
$E^{-}(r)=\widetilde{E^{-}}J_{l-1}\left(\mu r\right)$ and $E_{z}(r)=\widetilde{E_{z}}J_{l}\left(\mu r\right)$,
where $J_{l}\left(\mu r\right)$ is the $l-$th order Bessel function,
and $\widetilde{E^{+}}$, $\widetilde{E^{-}}$ and $\widetilde{E_{z}}$
are undetermined constants. Here, $\mu$ is a constant that can be
viewed as a special kind of wave number in the $r$-direction. All
three mode components have the same dispersion relation, 
\begin{eqnarray}
\omega^{2}=\omega_{p}^{2}+\left(k^{2}+\mu^{2}\right)c^{2},\label{eq:19}
\end{eqnarray}
which indicates that the three modes are degenerate states. However,
the divergence free condition, i.e., Eq.\,\eqref{eq:8}, puts on
a constraint on the $E^{+}(r)$, $E^{-}(r)$, and $E_{z}(r)$ components,
\begin{eqnarray}
\left(\frac{1}{r}\frac{\partial}{\partial r}r\!+\!\frac{l}{r}\right)E^{+}(r)+\!\left(\frac{1}{r}\frac{\partial}{\partial r}r\!-\!\frac{l}{r}\right)E^{-}(r)+\!i2kE_{z}(r)=\!0.\label{eq:18}
\end{eqnarray}
In terms of $\widetilde{E^{+}}$, $\widetilde{E^{-}}$ and $\widetilde{E_{z}}$,
it is 
\begin{eqnarray}
\mu\widetilde{E^{+}}-\mu\widetilde{E^{-}}+i2k\widetilde{E_{z}}=0.\label{eq:20}
\end{eqnarray}
For a given pair of $k$ and $\mu$, the mode has two degrees of freedom
or degeneracy.

The electromagnetic mode with OAM is localized around the wave axis,
and the amplitude envelope decays approximately as $1/\sqrt{r}$ for
large $r$. Because $J_{l}\left(0\right)=0$ when $l\neq0$, there
is no phase singularity of photon OAM at axis. The power density of
the mode maximized on a ring with a radius determined by the maximum
of the $z$-component of the momentum in Eq.\,\eqref{eq:P}.

Here, we discuss a special case with $\widetilde{E_{z}}=0$ and $\widetilde{E^{+}}=\widetilde{E^{-}}=\widetilde{E}$.
In this case, 
\begin{eqnarray}
E_{r}(r,\phi,z) & = & \frac{1}{2}\left[J_{l+1}\left(\mu r\right)+J_{l-1}\left(\mu r\right)\right]\widetilde{E}\mathrm{e}^{i\left(l\phi+kz-\omega t\right)},\label{eq:21}\\
E_{\phi}(r,\phi,z) & = & \frac{i}{2}\left[J_{l-1}\left(\mu r\right)-J_{l+1}\left(\mu r\right)\right]\widetilde{E}\mathrm{e}^{i\left(l\phi+kz-\omega t\right)}.\label{eq:22}
\end{eqnarray}
From Eqs.~(\ref{eq:21})-(\ref{eq:22}), the time averaged momentum
and angular momentum densities are 
\begin{align}
\left\langle \mathbf{P}\right\rangle  & =(\varepsilon_{0}/2)\mathrm{Re}\left(\mathbf{E}\times\mathbf{B}^{*}\right)=\frac{\varepsilon_{0}\widetilde{E}^{2}}{2\omega}\left[\frac{l}{r}J_{l}^{2}\mathbf{e}_{\phi}+\left(\frac{kl^{2}}{\mu^{2}r^{2}}J_{l}^{2}+kJ_{l}^{'2}\right)\mathbf{e}_{z}\right],\label{eq:P}\\
\left\langle \mathbf{M}\right\rangle  & =\mathbf{r}\times\left\langle \mathbf{P}\right\rangle =-\frac{\varepsilon_{0}\widetilde{E}^{2}}{2\omega}\left[\frac{zl}{r}J_{l}^{2}\mathbf{e}_{r}+\left(\frac{kl^{2}}{\mu^{2}r}J_{l}^{2}+krJ_{l}^{'2}\right)\mathbf{e}_{\phi}-lJ_{l}^{2}\mathbf{e}_{z}\right].
\end{align}
The radial component of $\left\langle \mathbf{M}\right\rangle \!$
and the azimuthal components of \textbf{$\left\langle \mathbf{P}\right\rangle $}
and $\left\langle \mathbf{M}\right\rangle $ are symmetric about the
axis, thus spatial average leaves only the $z$-components, which
shows that the eigen modes carry \textbf{$z$}-photon OAM. From Eqs.~(\ref{eq:2})--(\ref{eq:4}),
we can also find that the OAM of electrons is opposite to that of
ions, and the total OAM of electrons and ions is zero.

Different from the scalar paraxial solutions with slow varying envelope
approximation, the OAM eigen modes obtained above are rigorous analytical
solutions admitted by plasmas, which are orthogonal and complete.
It is not surprising to find the similarities and differences between
our solutions specified by Eqs.~(\ref{eq:21})-(\ref{eq:22}) and
the familiar Bessel modes, 
\begin{eqnarray}
E_{r}|_{\left\{ \cos(l\phi),\thinspace\sin(l\phi)\right\} } & = & \!\frac{l}{\mu r}J_{l}\left(\mu r\right)\left\{ \cos(l\phi),\thinspace\sin(l\phi)\right\} \widetilde{E}\mathrm{e}^{i\left(kz-\omega t\right)},\label{eq:26}\\
E_{\phi}|_{\left\{ \cos(l\phi),\thinspace\sin(l\phi)\right\} } & = & \!J_{l}^{'}\left(\mu r\right)\left\{ -\sin(l\phi),\thinspace\cos(l\phi)\right\} \widetilde{E}\mathrm{e}^{i\left(kz-\omega t\right)}.\label{eq:27}
\end{eqnarray}
Their radial dependencies are all expressed in terms of Bessel functions,
and they are both diffraction free, as there is no radial momentum
component. However, there are major differences. Equations~(\ref{eq:21})-(\ref{eq:22})
give an azimuthal phase distribution, which forms a helical wave front.
On the other hand, the familiar Bessel modes have two degenerate polarization
components, which have orthogonal azimuthal amplitude distributions.
Another important difference is that the Bessel modes carry no OAM,
which can be verified by direct calculation. Interestingly, an electromagnetic
mode with OAM can be constructed from two Bessel modes without OAM
as 
\begin{eqnarray}
A_{\beta}=A_{\beta}|_{\cos(l\phi)}+iA_{\beta}|_{\sin(l\phi)}.\label{eq:25}
\end{eqnarray}
Here, $A_{\beta}$ denotes mode components of the electromagnetic
modes with OAM obtained from Eq.~(\ref{eq:17}), and $A_{\beta}|_{\cos(l\phi)}$
and $A_{\beta}|_{\sin(l\phi)}$ are the degenerate Bessel modes without
OAM. The Euler formula $e^{l\phi}=\cos\left(l\phi\right)+i\sin\left(l\phi\right)$
realizes the conversion from orthogonal azimuthal amplitude distributions
to a topological charge. One may wonder why one OAM mode can be excited
by two modes without OAM? This effect can be attributed to the familiar
coherent interference. To wit, we have 
\begin{eqnarray}
\int\mathbf{r}\times\mathrm{Re}\left[\left(\mathbf{E}_{1}+i\mathbf{E}_{2}\right)\times\left(\mathbf{B}_{1}^{*}+i\mathbf{B}_{2}^{*}\right)\right]\mathrm{d}V=\int\mathbf{r}\times\mathrm{Im}\left(\mathbf{E}_{1}\times\mathbf{B}_{2}^{*}+\mathbf{E}_{2}\times\mathbf{B}_{1}^{*}\right)\mathrm{d}V,
\end{eqnarray}
where the superscripts 1 and 2 denote two degenerate states without
OAM in Eq.~(\ref{eq:25}). The cross product between electric and
magnetic fields of different modes leads to an azimuthal momentum
distribution, and thus a twisted beam. This phenomenon is similar
to the process that a circularly polarized wave with spin can be excited
by two linearly polarized waves without spin.

We now investigate the electrostatic modes with OAM. Substituting
Eq.~(\ref{eq:13}) into Eq.~(\ref{eq:9}), we obtain the electric
field eigen equations of electrostatic modes with OAM. These equations
can be written in a vector form as, 
\begin{eqnarray}
\frac{\omega^{2}-\omega_{p}^{2}}{c^{2}}\mathbf{E}\left(r\right)\mathrm{e}^{i\left(l\phi+kz-\omega{t}\right)}=0.\label{eq:29}
\end{eqnarray}
The dispersion relation obtained from Eq.~(\ref{eq:29}) is that
for plasma oscillation, i.e., $\omega=\omega_{p}$, which should not
be surprising. This mode can be viewed as a plasmon with OAM. The
components $E_{r}(r)$, $E_{\phi}(r)$ and $E_{z}(r)$ should satisfy
the rotation free condition, i.e., Eq.~(\ref{eq:10}), 
\begin{eqnarray}
\left(\begin{array}{ccc}
0 & -ik & i\frac{l}{r}\\
ik & 0 & -\frac{\partial}{\partial{r}}\\
-i\frac{l}{r} & \frac{1}{r}\frac{\partial}{\partial{r}}r & 0
\end{array}\right)\left(\begin{array}{c}
E_{r}(r)\\
E_{\phi}(r)\\
E_{z}(r)
\end{array}\right)=0.\label{eq:30}
\end{eqnarray}
Because the rank of the coefficient matrix in Eq.~(\ref{eq:30})
is 2, there are two constraints and one independent mode component.
In another word, the mode is non-degenerate. Solving Eq.~(\ref{eq:30}),
we obtain 
\begin{eqnarray}
E_{r}(r) & = & -i\frac{1}{k}\frac{\partial}{\partial r}E_{z}(r),\,\,\,E_{\phi}(r)=\frac{l}{kr}E_{z}(r),\label{eq:31}
\end{eqnarray}
where $E_{z}(r)$ is an arbitrary function of $r.$ However, in order
to avoid the photon OAM phase singularity, it should satisfy the following
conditions, 
\begin{eqnarray}
E_{z}\left(0\right)=0,\,\,\,\frac{E_{z}\left(r\right)}{r}|_{0}=0,\,\,\,\frac{\partial}{\partial r}E_{z}\left(r\right)|_{0}=0,\label{eq:33}
\end{eqnarray}
The polarization properties described by Eq.~(\ref{eq:31}) show
that for the electrostatic mode with OAM, the electrical field is
not parallel to the space-time averaged propagation axis, which is
in the $z$-direction. This situation is similar to the fact that
for the electromagnetic mode with OAM, the electrical field is not
perpendicular to the space-time averaged propagation axis.

We note that the electrostatic mode, or the plasmon, with OAM is a
non-propagating oscillation under the cold plasma approximation. We
now investigate finite temperature effects, one of which is the formation
of a new propagating electrostatic mode with OAM, i.e., phonon with
OAM. When the finite temperature is considered, Eq.~(\ref{eq:2})
should be modified as 
\begin{eqnarray}
\frac{\partial}{\partial t}\mathbf{V}_{\alpha}=-\frac{1}{n_{0\alpha}m_{\alpha}}\bigtriangledown p_{1\alpha}+\frac{q_{\alpha}}{m_{\alpha}}\mathbf{E},\label{eq:34}
\end{eqnarray}
where the thermal pressures $p_{1\alpha}$ satisfies the polytropic
law $p_{1\alpha}/p_{0\alpha}=\gamma_{\alpha}n_{1\alpha}/n_{0\alpha}$.
The thermal velocities for electron and ion are defined as $V_{T\alpha}=\sqrt{\gamma_{\alpha}p_{0\alpha}/n_{0\alpha}m_{\alpha}}=\sqrt{\gamma_{\alpha}k_{B}T_{0\alpha}/m_{\alpha}}$,
where the $\gamma_{\alpha}$ is the polytropic index. Substituting
Eq.~(\ref{eq:1}) into Eq.~(\ref{eq:34}), we obtain, 
\begin{eqnarray}
-\omega^{2}\mathbf{V}_{\alpha}=\frac{\gamma_{\alpha}p_{0\alpha}}{n_{0\alpha}m_{\alpha}}\bigtriangledown\bigtriangledown\cdot\mathbf{V}_{\alpha}+\frac{Q_{\alpha}}{m_{\alpha}}\mathbf{E}.\label{eq:35}
\end{eqnarray}

For the electromagnetic modes, Eqs.~(\ref{eq:7}) and (\ref{eq:35})
lead to Eq.~(\ref{eq:14}), which means that there is no thermal
correction for the electromagnetic modes with OAM. The thermal effect
on the electrostatic modes in more interesting. It produces phonons
with OAM in plasmas. With finite temperature, it is more convenient
to derive the eigen system using the velocity components. Equations
(\ref{eq:9}) and (\ref{eq:35}) lead to 
\begin{eqnarray}
\left(\begin{array}{cccccc}
S_{e} & -iD & 0 & U_{e} & 0 & 0\\
iD & S_{e} & 0 & 0 & U_{e} & 0\\
0 & 0 & P_{e} & 0 & 0 & U_{e}\\
U_{i} & 0 & 0 & S_{i} & -iD & 0\\
0 & U_{i} & 0 & iD & S_{i} & 0\\
0 & 0 & U_{i} & 0 & 0 & P_{i}
\end{array}\right)\left(\begin{array}{c}
V_{re}(r)\\
V_{\phi e}(r)\\
V_{ze}(r)\\
V_{ri}(r)\\
V_{\phi i}(r)\\
V_{zi}(r)
\end{array}\right)=0,\label{eq:36}
\end{eqnarray}
for $\mathbf{V}_{\alpha}=\sum_{\beta}V_{\alpha\beta}\left(r,\phi,z\right)\mathbf{e}_{\beta}=\sum_{\beta}V_{\alpha\beta}\left(r\right)\mathrm{e}^{i\left(l\phi+kz-\omega t\right)}\mathbf{e}_{\beta}$.
Here, $U_{\alpha}=\omega_{p\alpha}^{2}/V_{T\alpha}^{2}$, $D$ was
defined after Eq.\,\eqref{eq:14}, and $S_{\alpha}$ and $P_{\alpha}$
have similar forms as Eqs.~(\ref{eq:15}) and (\ref{eq:16}), except
that $c^{2}$ is replaced by $V_{T\alpha}^{2}$ and $\omega_{p}^{2}$
is replaced by $\omega_{p\alpha}^{2}$, respectively.

Defining new field components $V_{\alpha}^{\pm}(r)=V_{r\alpha}(r)\pm iV_{\phi\alpha}(r)$,
we can rewrite Eq.~(\ref{eq:36}) in the principal axis system as

\begin{eqnarray}
\left(\begin{array}{cccccc}
S_{e}\!-\!D & 0 & 0 & U_{e} & 0 & 0\\
0 & S_{e}\!+\!D & 0 & 0 & U_{e} & 0\\
0 & 0 & P_{e} & 0 & 0 & U_{e}\\
U_{i} & 0 & 0 & S_{i}\!-\!D & 0 & 0\\
0 & U_{i} & 0 & 0 & S_{i}\!+\!D & 0\\
0 & 0 & U_{i} & 0 & 0 & P_{i}
\end{array}\right)\left(\begin{array}{c}
V_{e}^{+}(r)\\
V_{e}^{-}(r)\\
V_{ze}(r)\\
V_{i}^{+}(r)\\
V_{i}^{-}(r)\\
V_{zi}(r)
\end{array}\right)\!=\!0.\label{eq:37}
\end{eqnarray}
It shows that the eigen system of the electrostatic mode with OAM
in a warm plasma consists of three decoupled subsystems $\left(V_{e}^{+},V_{i}^{+}\right)$,
$\left(V_{e}^{-},V_{i}^{-}\right)$ and $\left(V_{ze},V_{zi}\right)$.
These subsystems have same dispersion relations, representing three
degenerate states with different polarizations. For the subsystem
$\left(V_{ze},V_{zi}\right)$, for example, with $V_{z\alpha}(r)=\widetilde{V_{z\alpha}}J_{l}\left(\mu r\right)\mathrm{e}^{i\left(l\phi+kz-\omega t\right)}$,
the eigen equations are, 
\begin{eqnarray}
\left(\begin{array}{cc}
\frac{\omega^{2}\!-\!\omega_{pe}^{2}}{V_{Te}^{2}}\!-\!\mu^{2}\!-\!k^{2} & \frac{\omega_{pe}^{2}}{V_{Te}^{2}}\\
\frac{\omega_{pi}^{2}}{V_{Ti}^{2}} & \frac{\omega^{2}\!-\!\omega_{pi}^{2}}{V_{Ti}^{2}}\!-\!\mu^{2}\!-\!k^{2}
\end{array}\right)\left(\begin{array}{c}
\widetilde{V_{ze}}\\
\widetilde{V_{zi}}
\end{array}\right)\!=\!0.\label{eq:38}
\end{eqnarray}
The dispersion relation given by Eq.~(\ref{eq:38}) is 
\begin{widetext}
\begin{eqnarray}
\omega^{2}=\frac{1}{2}\left[\omega_{p}^{2}+V_{T}^{2}\left(\mu^{2}+k^{2}\right)\right]\left[1\pm\sqrt{1-\frac{4\Omega^{2}}{\left[\omega_{p}^{2}+V_{T}^{2}\left(\mu^{2}+k^{2}\right)\right]^{2}}}\right]\thinspace,\label{eq:39}\\
\Omega^{2}\equiv\left(\mu^{2}+k^{2}\right)V_{Te}^{2}V_{Ti}^{2}+\left(\mu^{2}+k^{2}\right)V_{Ti}^{2}\omega_{pe}^{2}+\left(\mu^{2}+k^{2}\right)V_{Te}^{2}\omega_{pi}^{2}\thinspace,
\end{eqnarray}
where $V_{T}=\sqrt{V_{Te}^{2}+V_{Ti}^{2}}$. The polarization relation
between electron and ion velocities is specified by 
\begin{gather}
V_{zi}=KV_{ze},\thinspace V_{i}^{+}=KV_{e}^{+},\thinspace V_{i}^{-}=KV_{e}^{-},
\end{gather}
where 
\begin{equation}
K\equiv\left(\frac{V_{Te}^{2}}{\omega_{pe}^{2}}\right)\left(\mu^{2}+k^{2}-\frac{\omega^{2}-\omega_{pe}^{2}}{V_{Te}^{2}}\right)=\left(\frac{\omega_{pi}^{2}}{V_{Ti}^{2}}\right)\left(\mu^{2}+k^{2}-\frac{\omega^{2}-\omega_{pi}^{2}}{V_{Ti}^{2}}\right).\label{eq:K}
\end{equation}
In Eq.\,\eqref{eq:K}, the second equal sign is another, probably
more transparent, way to write the dispersion relation \eqref{eq:39}.
In most cases $V_{Ti}<<V_{Te}$, and the two branches of Eq.~(\ref{eq:39})
can be simplified as 
\begin{eqnarray}
\omega^{2} & \!\approx\! & \omega_{p}^{2}\!+\!V_{T}^{2}\left(\mu^{2}\!+\!k^{2}\right),\label{eq:40}\\
\frac{\omega^{2}}{\mu^{2}\!+\!k^{2}} & \!\approx\! & \frac{\left(\mu^{2}\!+\!k^{2}\right)^{2}V_{Te}^{2}V_{Ti}^{2}\!+\!V_{Ti}^{2}\omega_{pe}^{2}\!+\!V_{Te}^{2}\omega_{pi}^{2}}{\omega_{p}^{2}\!+\!V_{T}^{2}\left(\mu^{2}\!+\!k^{2}\right)}.\label{eq:41}
\end{eqnarray}
Equation~(\ref{eq:40}) describes the Langmuir wave with OAM, which
is a propagating plasmon with OAM. Equation~(\ref{eq:41}) is the
dispersion relation for the electrostatic mode which vanishes in the
cold plasma limit. It is the low frequency ion acoustic wave with
OAM. It can be viewed as a phonon with OAM. The OAM of the modes can
be calculated from the eigen structure. Because an electrostatic mode
does not carry electromagnetic momentum density \cite{Qin14-043102},
the mode contains only kinetic momentum density of the particles.
The fist order density field of the mode is 
\begin{eqnarray}
n_{1\alpha}=\frac{n_{0\alpha}\widetilde{V_{z\alpha}}}{\omega}\left(-\frac{\mu^{2}}{k}J_{l}^{''}\left(\mu r\right)-\frac{\mu}{kr}J_{l}^{'}\left(\mu r\right)+\frac{k^{2}r^{2}+l^{2}}{kr^{2}}J_{l}\left(\mu r\right)\right)\mathrm{e}^{i\left(l\phi+kz-\omega t\right)}.
\end{eqnarray}
The time averaged momentum density and angular momentum density of
plasma components are, 
\begin{eqnarray}
\left<\mathbf{P}_{\alpha}\right> & = & \mathrm{Re}\left(m_{\alpha}n_{1\alpha}^{*}\mathbf{V}_{\alpha}\right)=\frac{l}{kr}R_{\alpha}\left(r\right)\mathbf{e}_{\phi}+R_{\alpha}\left(r\right)\mathbf{e}_{z},
\end{eqnarray}
\begin{eqnarray}
\left<\mathbf{M}_{\alpha}\right> & = & \mathbf{r}\times\left<\mathbf{P}_{\alpha}\right>=-\frac{zl}{kr}R_{\alpha}\left(r\right)\mathbf{e}_{r}-rR_{\alpha}\left(r\right)\mathbf{e}_{\phi}+\frac{l}{k}R_{\alpha}\left(r\right)\mathbf{e}_{z},
\end{eqnarray}
where 
\begin{eqnarray}
R_{\alpha}\left(r\right)\equiv\frac{m_{\alpha}n_{0\alpha}\widetilde{V_{z\alpha}}^{2}}{\omega}J_{l}\left(\mu r\right)\left(-\frac{\mu^{2}}{k}J_{l}^{''}\left(\mu r\right)-\frac{\mu}{kr}J_{l}^{'}\left(\mu r\right)+\frac{k^{2}r^{2}+l^{2}}{kr^{2}}J_{l}\left(\mu r\right)\right).
\end{eqnarray}
The radial component of $\left<\mathbf{M}_{\alpha}\right>$ and the
azimuthal components of \textbf{$\left<\mathbf{P}_{\alpha}\right>$}
and $\left<\mathbf{M}_{\alpha}\right>$ are symmetric about the axis,
and integration over space leaves only the z-components, which show
that the plasma components carry z-plasmon or z-phonon OAM. Furthermore,
when $m_{e}n_{0e}+m_{i}n_{0i}K^{2}\neq0$, the mode contains a global
OAM. 
\end{widetext}

The unique properties of the OAM phontons, phonons, and plasmons discussed
above enable important potential applications in plasma physics and
accelerator physics. As an intrinsic characteristic of the OAM beam,
the highly localized power density off the propagation axis can be
an effective tool for delivering focused heating and acceleration
power. It is also a potential plasma diagnostic technique. The OAM
states can be modulated by inhomogeneous and anisotropic structures
in plasmas, such as density and magnetic field fluctuations. By injecting
a OAM beam and detecting the OAM scattering spectrum, we can infer
statistical properties of the fluctuation in the plasma. For application
in accelerator physics, if electromagnetic modes with OAM are introduced
as accelerating field structures, charged particle beams will be driven
by the OAM of the modes to rotate. Rotating particle beams are more
stable for applications where high beam intensity is required.

In this work, electromagnetic and electrostatic waves with OAM in
unmagnetized homogeneous plasmas are systematically studied. Exact
OAM eigen modes are derived, which are different from approximate
solutions in scalar paraxial optics with slow varying envelopes. Three
classes of OAM modes are discovered: photons, phonons, and plamsons,
which correspond to the electromagnetic, ion acoustic, and Langmuir
waves. The modes have azimuthal phase distribution and Bessel-type
of radial dependency. It is found that the electromagnetic mode with
OAM can be excited by two familiar Bessel modes without OAM. For the
phonons and plasmons, the OAM are carried by the electrons and ions.
The OAM modes in plasmas and their characteristics can be explored
for various potential applications. Further studies of the propagation
properties of the modes with OAM and their interactions with plasmas
are expected to reveal new physics previous unknown. 
\begin{acknowledgments}
This research is supported by the National Natural Science Foundation
of China (NSFC-51477182, 11505186, 11575185, 11575186) and ITER-China
Program (2015GB111003, 2014GB124005). 
\end{acknowledgments}

%\bibliographystyle{apsrev4-1}
%\bibliography{QchenOAM}
\providecommand{\noopsort}[1]{}\providecommand{\singleletter}[1]{#1}%

\end{document}